\documentclass{article}
\usepackage{amsmath,amssymb}
\usepackage{graphicx}%
\newcommand{\Fig}[3]{%
\begin{center}
\parbox{#2cm}{%
\refstepcounter{figure}\includegraphics[width=#2cm]{#1}\\[12pt] \noindent {\bf Fig. \thefigure.}\quad
#3}\end{center}}

\newcommand{\TwoFig}[4]{%
\begin{center}
\begin{tabular}{lr}
\parbox{7.7cm}{\includegraphics[width=7.7cm]{#1}}  & \parbox{7.7cm}{\includegraphics[width=7.7cm]{#2}} \\
\parbox{7.7cm}{\refstepcounter{figure}{\bf Fig. \thefigure.}\quad #3} & \parbox{7.7cm}{\refstepcounter{figure}{\bf Fig. \thefigure.}\quad #4}\\
\end{tabular}
\end{center}
}
\begin{document}

\begin{center}
{\bf \large 
A qualitative and numerical analysis of cosmological models\\[4pt] 
based on assymetric scalar doublet:\\[4pt]  classical + phantom scalar field.\\[4pt]  I. A case of minimally interacting scalar fields:  \\[4pt]the qualitative analysis.} \\[8pt]
Yu.\,G.~Ignat'ev\\
N.I. Lobachevsky Institute of Mathematics and Mechanics, Kazan Federal University, \\ Kremleovskaya str., 35, Kazan, 420008, Russia
\end{center}

\begin{abstract}
The paper provides a qualitative and numerical analysis-based investigation of cosmological models founded on an asymmetrical scalar doublet comprising of a classical and a phantom scalar fields. Presence of a phantom scalar field allows one to consider also classical scalar fields with attraction of like-charged particles which significantly extends a diversity of cosmological models' behaviours. It is shown that a cosmological model based on an asymmetric scalar doublet in the case of minimal interaction has 9 singular points 2 of which are attractive and the rest are non-stable saddle ones. It is also shown that a presence of even essentially weak phantom field significantly changes the dynamics of a cosmological model.

{\bf keyword} phantom scalar fields, %
scalar particle interaction, asymmetric scalar doublet, cosmological models, qualitative analysis, numerical simulation\\
{\bf PACS}: 04.20.Cv, 98.80.Cq, 96.50.S  52.27.Ny
\end{abstract}

This work was funded by the subsidy allocated to Kazan Federal University for the state assignment in the sphere of scientific activities.

\section{Introduction}
Standard cosmological models (SCM) \footnote{see e.g. \cite{Gorb_Rubak}}, based on a classical scalar field, were investigated by methods of the qualitative analysis of dynamic systems in \cite{Belinsky}, \cite{Zeld},  \cite{Zhur_01}, \cite{Mex1},\cite{Mex2} (see also \cite{Bron}). In paper  \cite{Zhur_01} it was also investigated a two-component cosmological model with minimum interaction (see also \cite{Zhur})\footnote{see also \cite{Mex2}}. In the Author's work \cite{Ignat_16_1_stfi} the qualitative and also numerical analysis of the standard cosmological model based on classical scalar field was over again carried out; it was done reducing the problem to investigation of the dynamic system on a 2-dimensional phase plane $\{\Phi,\dot{\Phi}\}$. A microscopic oscillating character of the invariant cosmological acceleration at late stages of expansion was also shown for this case. The results were generalized to cosmological models with a $\Lambda$ - term \cite{Ignat_16_2_stfi}, \cite{Ignat_17_1_GC}\footnote{see also \cite{Ignat_16_4_Mono}},  and the Authors managed to confirm a conversation of the oscillating character of the invariant cosmological acceleration at significantly small values of the cosmological term. Moreover,  averaging of the cosmological acceleration by microscopic oscillations in the latest articles helped to illustrate a possibility of the macroscopic acceleration's escape to a non-relativistic mode at late stages of the early Universe \footnote{i.e., at stages with scalar field's dominance}.
The suggested in the above cited papers investigation method was used in V.M.Zhiravlev's work \cite{Zhur_16}. It was applied to a two-component system  <<scalar field + liquid>> with an arbitrary potential function $V(\phi)$\footnote{in particular, for the Higgs potential.}.
\par On a formal level, phantom fields were introduced into gravitation in the capacity of one of the possible models of a scalar field in 1983 in the Author's work \cite{Ignat83_1}. Phantom fields in this and later papers (see e.g., \cite{Ignat_Kuz84}, \cite{Ignat_Mif06}) were classified as scalar fields with an attraction of like-charged particles and were distinguished by the factor $\epsilon=-1$ in the energy-momentum tensor of a scalar field. Let us notice that phantom fields in respect to wormholes and the so-called black universes were considered in papers  \cite{Bron1}, \cite{Bron2}\footnote{Here it is necessary to mention the fact that a concept of being <<phantom>> in our papers and cited above is somewhat different, coinciding in its essence in the case of a phantom field with repulsion.}. let us also notice that a single classical scalar field with attraction can not exist since it corresponds to purely negative total energy. Such a field can theoretically exit only in a multiplet with other fields.
\par Thus, phantom scalar fields with attraction of like-charged particles are corresponded by a negative kinetic term in the energy-momentum tensor, while phantom fields with repulsion are corresponded by a positive kinetic term. However, in both cases they are corresponded by positive signs of the kinetic and massive terms in the Klein-Gordon equation. Corresponding solution for single-isolated charge do not take form of the Yukawa potential but rather they take form of the solutions of Lifshitz scalar perturbations' equations for spherical symmetry ($\sin kr/r$) \cite{Ignat_12_3_Iz}.
\par The nonminimal theory of scalar interaction was being sequentially developed on the basis of the fundamental charge's concept for both classical and phantom scalar fields \cite{Ignat_12_1_Iz}, \cite{Ignat_12_2_Iz}, \cite{Ignat_12_3_Iz}, \cite{Ignat_13_Mono}. In particular, these works have revealed certain peculiarities of phantom fields, for instance, peculiarities of interparicle interaction. Later, these researches were elaborated for propagation of the theory of scalar fields including phantom ones to a sector of non-negative masses of particles, degenerated Fermi - systems, conformal invariant interactions etc. \cite{Ignat_14_1_stfi}, \cite{Ignat_Dima14_2_GC}, \cite{Ignat_Agaf_Dima14_3_GC}, \cite{Ignat_15_1_GC}, \cite{Ignat_Agaf15_2_GC}. Mathematical models of scalar fields being constructed in this way, were applied to investigation of the cosmological evolution of systems of interacting particles and scalar fields of both classical and phantom types \cite{Ignat_Mih15_2_Iz}, \cite{Ignat_Agaf_Mih_Dima15_3_AST}, \cite{Ignat_Agaf_16_3}. These researches revealed unique properties of cosmological evolution of plasma with interparticle phantom scalar interaction, such as the existence of giant bursts of the cosmological acceleration, presence of a plateau with constant acceleration, and other anomalies which markedly differ the behaviour of cosmological models with phantom scalar field and the behaviour of those with classical scalar field. In particular, a classification of the behaviour types of cosmological models with interparticle phantom scalar field was carried out in works \cite{Ignat_Agaf_16_3} -- \cite{Ignat_Sasha_G&G}. In the result, 4 essentially different models were distinguished. In the same works it was also pointed out a possibility of Bose condensation of nonrelativistic scalar charged fermions in conditions of strong growth of a scalar field's potential and consideration of this condensate in the capacity of a dark matter's component. It is important to highlight the next circumstance: the values of cosmological acceleration that are greater than 1 are achievable in the course of cosmological evolution. Such values, in accordance with standard classification correspond exactly to phantom state of matter.

\par Cited above researches show the necessity to investigate phantom scalar fields in the capacity of possible basis of the cosmological model of the early Universe.  In works \cite{Ignat_16_5_Iz} -- \cite{Ignat_Agaf_2017_2_GC} a preliminary qualitative analysis of the cosologcal model based on a phantom scalar field with self-action has been carried out. In the given paper, we develop and specify the results of investigations of cosmological models based on classical and phantom scalar fields. In contrast to works \cite{Ignat_14_1_stfi} -- \cite{Ignat_Agaf_Mih_Dima15_3_AST} we do not take into account a contribution of the dark matter, i.e. we consider free classical and phantom fields without a source and investigate a combined system comprising of a couple of scalar fields, classical and phantom ones. Further, we will be calling a system comprising of two scalar fields, classical, $\Phi$, and phantom, $\phi$, an \emph{asymmetrical scalar doublet.} In this paper we consider a most simple case of free scalar fields interacting between each other only by means of gravitation and also confine ourselves mainly to qualitative research of such dynamic system leaving the results of numerical simulation and the investigation of cosmological evolution of asymmetric scalar doublet with fields interacting between themselves to further parts of this work. Let us notice that even such a simplified task turns to be much more complicated than those considered in the cited above papers since it is reduced to research of 4-dimensional dynamic system in contrast to two-dimensional systems investigated earlier. Therefore in this paper we confine ourselves to classification of the system's singular points and represent certain results of numerical simulation of the dynamical system in the neighborhood of its null singular point.

\section{Main Relations of the Cosmological Model Based on Asymmetric Scalar Doublet}
\subsection{The Field Equations}
Let us write down the Lagrangian function of the scalar doublet comprising of classical and phantom scalar fields with a self-action in the Higgs form and minimal interaction between themselves in the following form:
\begin{equation}\label{Lagrange0}
L=\frac{\epsilon_1}{8\pi}\left(g^{ik}\Phi_{,i}\Phi_{,k}-2V(\Phi)\right)+\frac{\epsilon'_1}{8\pi}\left(g^{ik}\phi_{,i}\phi_{,k}-2v(\phi)\right),
\end{equation}
where
\begin{eqnarray}
\label{V(phi)}
V(\Phi)= -\frac{\alpha}{4}\left(\Phi^2+\frac{m^2}{\alpha}\right)^2;\\
v(\phi)= -\frac{\beta}{4}\left(\phi^2-\frac{\mathfrak{m}^2}{\beta}\right)^2
\end{eqnarray}
is a potential Higgs energy\footnote{It is necessary to remember that a scalar field's potential energy in accordance with (\ref{Lagrange0}) actually equals to $\epsilon_1 V(\Phi)$} of the corresponding scalar fields where $\alpha $ and $\beta$ are the constants of their self-action, $m$ and $\mathfrak{m}$ are their masses of the quanta; for a field with a repulsion of like-charged particles it is $\epsilon_1,\epsilon'_1=1$, for a field with attraction of like-charged particles it is $\epsilon_1,\epsilon'_1=-1$.

The scalar field's energy-momentum tensor relative to the Lagrangian function (\ref{Lagrange0}) takes the standard form:
\begin{equation}\label{T_ik0}
T_{ik}=\frac{\epsilon_1}{8\pi}\bigl(2\Phi_{,i}\Phi_{,k}-g_{ik}\Phi_{,j}\Phi^{,j}+2V(\Phi)g_{ik}\bigr)+
\frac{\epsilon'_1}{8\pi}\bigl(2\phi_{,i}\phi_{,k}-g_{ik}\phi_{,j}\phi^{,j}+2v(\phi)g_{ik}\bigr).
\end{equation}
Variation of the Lagrangian function (\ref{Lagrange0}) leads to the field equations:
\begin{eqnarray} \label{EqField0_Phi}
\square \Phi +V'(\Phi) =0;\\
\label{EqField0_phi}
\square \phi +v'(\phi) =0.
\end{eqnarray}
Since we can add an arbitrary constant to the Lagrangian function\footnote{which leads to renormalization of the cosmological constant}, further we omit corresponding constants in the potential functions where it leads to simplifications. Such renormalization returns us to the initial Lagrangian function of the scalar field with self-action which we had in \cite{Ignat_16_5_Iz} -- \cite{Ignat_Agaf_2017_2_GC}, and which we will use further (see \cite{Ignat83_1}):
\begin{equation} \label{Lagrange}
L=\frac{\epsilon_1}{8\pi } \left(g^{ik} \Phi _{,i} \Phi _{,k} -m^{2} \Phi ^{2} +\frac{\alpha }{2} \Phi ^{4} \right)+
\frac{\epsilon'_1}{8\pi } \left(g^{ik} \phi _{,i} \phi _{,k} + \mathfrak{m}^{2} \phi ^{2} +\frac{\beta}{2} \phi ^{4} \right),
\end{equation}
The energy-momentum tensor relative to the Lagrangian function (\ref{Lagrange}) is equal to
\begin{eqnarray}\label{T_ik}
T_{ik}=\frac{\epsilon_1}{8\pi}\bigl(2\Phi_{,i}\Phi_{,k}-g_{ik}\Phi_{,j}\Phi^{,j}+g_{ik} m^2\Phi^2-g_{ik}\frac{\alpha}{2}\Phi^4\bigr)
\nonumber\\
+\frac{\epsilon'_1}{8\pi}\bigl(2\phi_{,i}\phi_{,k}-g_{ik}\phi_{,j}\phi^{,j}-g_{ik}\mathfrak{m}^2\phi^2-g_{ik}\frac{\beta}{2}\phi^4\bigr).
\end{eqnarray}
Let us obtain the equations of free classical and phantom fields carrying out a standard variational procedure over the Lagrangian function in form of (\ref{Lagrange}):
\begin{eqnarray} \label{EqField1_c}
\Box\Phi+m^2_*\Phi=0;\\
\label{EqField1_f}
\Box\phi-\mathfrak{m}^2_*\phi=0,
\end{eqnarray}
where $m_*, \mathfrak{m}_c$ are the effective masses of scalar bosons
\begin{eqnarray}\label{m_*}
m^2_* =& \epsilon_2 m^2-\alpha\Phi^2;\\
\mathfrak{m}^2_* =& \epsilon_2 \mathfrak{m}^2+\beta\phi^2,
\end{eqnarray}
which theoretically can be imaginary values.

Let us write also the Einstein equations with a cosmological term\footnote{We use Planck system of units:
$G=c=\hbar =1$; Ricci tensor is obtained by means of convolution of first and four indices $R_{ik}=R^j_{~ikj}$; the metrics has signature $(-1,-1,-1,+1)$.}
\begin{equation} \label{EqEinstein0}
R^{ik} -\frac{1}{2} Rg^{ik} =\lambda g^{ik} +8\pi T^{ik},
\end{equation}
where $\lambda\geq 0$ is the cosmological constant.
\subsection{The Equations of the Cosmological Model}
Let us write out a self-consistent system of equations of the cosmological model (\ref{EqField1_c}), (\ref{EqField1_f}), (\ref{EqEinstein0}), based on the free asymmetric scalar doublet and space - flat Friedmann metrics
\begin{equation}\label{metric}
ds^{2} =dt^{2} -a^{2} (t)(dx^{2} +dy^{2} +dz^{2}),
\end{equation}
assuming $\Phi=\Phi(t)$, $\phi=\phi(t)$.

In this case the energy -- momentum tensor (\ref{T_ik}) has a structure of energy -- momentum tensor of ideal flux with summary energy density $\varepsilon$ and pressure $p$:
\begin{equation}\label{e,p}
\varepsilon(t)=\varepsilon_c+\varepsilon_f;\quad p(t)=p_c+p_f:
\end{equation}
\begin{eqnarray} \label{e_c,p_c}
\varepsilon_c =\frac{\epsilon_1}{8\pi } \left(\dot{\Phi }^{2} +m^{2}\Phi^{2} -\frac{\alpha }{2} \Phi ^{4} \right); &
\displaystyle p_c=\frac{\epsilon_1}{8\pi } \left(\dot{\Phi }^{2} -m^{2}\Phi ^{2} +\frac{\alpha }{2} \Phi ^{4} \right);\\
\label{e_f,p_f}
\varepsilon_f =\frac{\epsilon'_1}{8\pi } \left(\dot{\phi }^{2} - \mathfrak{m}^{2}\phi^{2} -\frac{\beta }{2} \phi ^{4} \right); &
\displaystyle p_c=\frac{\epsilon'_1}{8\pi } \left(\dot{\phi }^{2} + \mathfrak{m}^{2}\phi ^{2} +\frac{\beta }{2} \phi ^{4} \right),\\
\end{eqnarray}
where $\dot{f}\equiv df/dt$. Herewith the following relation is valid:
\begin{equation}\label{e+p}
\varepsilon +p=\frac{\epsilon_1}{4\pi }\dot{\Phi}^{2}+\frac{\epsilon'_1}{4\pi }\dot{\phi}^{2} .
\end{equation}

 Mentioned system comprises of a single Einstein equation
\begin{equation}\label{EqEinstein}
3\frac{\dot{a}^{2}}{a^{2}}\equiv 3H^2 =\epsilon_1\left(\dot{\Phi }^{2} + m^{2}\Phi^{2} -\frac{\alpha }{2} \Phi ^{4} \right) +
\epsilon'_1\left(\dot{\phi }^{2} - \mathfrak{m}^{2}\phi^{2} -\frac{\beta }{2} \phi ^{4} \right)+\lambda
\end{equation}
and two equations of the scalar field:
\begin{eqnarray}\label{Eq_fild_c}
\ddot{\Phi }+3\frac{\dot{a}}{a} \dot{\Phi }+m_{*}^{2} \Phi =0,\\
\label{Eq_fild_f}
\ddot{\phi }+3\frac{\dot{a}}{a} \dot{\phi }-\mathfrak{m}_{*}^{2} \phi =0.
\end{eqnarray}

In the given paper we consider a classical field with repulsion of like-charged particles assuming $\varepsilon_1=+1$. In this case the system of equations (\ref{EqEinstein}), (\ref{Eq_fild_c}), (\ref{Eq_fild_f}) takes its final form:
\begin{equation}\label{EqEinstein1}
3\frac{\dot{a}^{2}}{a^{2}} =\dot{\Phi }^{2} +m^{2}\Phi^{2} -\frac{\alpha }{2} \Phi ^{4} +e'_1\left(\dot{\phi }^{2} -\mathfrak{m}^{2}\phi^{2} -
\frac{\beta }{2} \phi ^{4}\right)+ \lambda,
\end{equation}
\begin{equation}\label{Eq_fild_c1}
\ddot{\Phi }+3\frac{\dot{a}}{a} \dot{\Phi }+m^{2} \Phi -\alpha\Phi^3=0,
\end{equation}
\begin{equation}\label{Eq_fild_f1}
\ddot{\phi }+3\frac{\dot{a}}{a} \dot{\phi }-\mathfrak{m}^{2} \phi -\beta\phi^3=0.
\end{equation}
\par Further we will require the following values of two kinematic functions of the Friedman Universe:
\begin{equation} \label{GrindEQ__10_}
H(t)=\frac{\dot{a}}{a} \ge 0;{\rm \; \; }\Omega (t)=\frac{a\ddot{a}}{\dot{a}^{2} } \equiv 1+\frac{\dot{H}}{H^{2} }
\end{equation}
the Hubble constant $H(t)$ and the invariant cosmological acceleration $\Omega(t)$, which is an invariant and expressed by means of a \emph{barotrope coefficient} $\varkappa=p/\varepsilon$ in the following way:
\begin{equation} \label{kappa}
\Omega= -\frac{1}{2}(1+3\varkappa).
\end{equation}
\section{The Qualitative Analysis}
\subsection{Reducing the System of Equations to the Canonical Form}
Making use of possibility to express the Hubble constant from  the Einstein equations (\ref{EqEinstein}) through functions $\Phi,\dot{\Phi}$, proceeding to dimensionless \emph{Compton time}:
\begin{equation}\label{tau}
mt=\tau;\quad (m\not\equiv0)
\end{equation}
and carrying out a standard change of variables $\Phi'=Z(\tau)$, $\phi'=z$, ($f'\equiv df/d\tau$), we reduce the Einstein equation (\ref{EqEinstein}) to dimensionless form:
\begin{equation}\label{Hm}
H'\ \!\!^2_m=\frac{1}{3}\left[Z^{2} +\Phi^{2} -\frac{\alpha_m}{2} \Phi ^{4} +
\epsilon'_1\left(z^{2} - \mu^{2}\phi^{2} -\frac{\beta_m}{2} \phi ^{4} \right)+\lambda_m\right],
\end{equation}
and the field equations (\ref{Eq_fild_c}), (\ref{Eq_fild_f}) - to the form of canonical autonomous system of ordinary differential equations in the 4-dimensional phase space $\mathbb{R}_4: \{\Phi,Z,\phi,z\}$:
\begin{eqnarray} \label{DynSys}
\Phi ' &=& Z;\nonumber\\
Z' &=& \displaystyle-\sqrt{3}Z \sqrt{Z^{2} +\Phi^{2} -\frac{\alpha_m}{2} \Phi ^{4} +
\epsilon'_1\left(z^{2} - \mu^{2}\phi^{2} -\frac{\beta_m}{2} \phi ^{4} \right)+\lambda_m} \nonumber - \Phi +\alpha _{m} \Phi ^{3} ,\\
\phi ' &=& z;\nonumber\\
z' &=& \displaystyle-\sqrt{3}z \sqrt{Z^{2} +\Phi^{2} -\frac{\alpha_m}{2} \Phi ^{4} +
\epsilon'_1\left(z^{2} - \mu^{2}\phi^{2} -\frac{\beta_m}{2} \phi ^{4} \right)+\lambda_m} + \phi +\beta _{m} \phi ^{3} ,
\end{eqnarray}
where the next dimensionless parameters are introduced:
\[\lambda _{m} \equiv \frac{\lambda }{m^{2}};\quad \alpha _{m} \equiv \frac{\alpha }{m^{2} };\quad \beta _{m} \equiv \frac{\beta }{m^2};
\quad \mu \equiv \frac{\mathfrak{m}}{m} . \]
Here it is:
\begin{equation} \label{GrindEQ__12_}
\frac{a'}{a}\equiv \Lambda'=H_m\equiv \frac{H}{m};\quad \Omega =\frac{aa''}{a'^{2} } \equiv 1+\frac{h'}{h^{2} } ,
\end{equation}
and
\begin{equation}\label{Lambda}
\Lambda=\ln a(\tau).
\end{equation}
Let us notice that in this notation all the values of the problem $\Phi, Z, H_m, \alpha_m, \beta_m, \mu, \Omega, \tau$ are dimensionless and time $\tau$ is measured in Compton scale with respect to a classical scalar field.

Thus, we have an autonomous 2-dimensional dynamic system in the 4-dimensional phase space $\mathbb{R}_4: \{\Phi,Z,\phi,z\}$:. To reduce it to standard notation of the qualitative theory of differential equations (see e.g.,  \cite{Bogoyav})
\begin{equation}\label{standart_sys}
\frac{dx_i}{d\tau}=F_i(x_1,\ldots,x_n),\quad i=\overline{1,n}
\end{equation}
and to simplify the writing let us accept the following denotations:
\begin{eqnarray}\label{GrindEQ__13_}
\Phi =x;\;& \displaystyle\quad \phi=y; \qquad F_1\equiv P=Z; \qquad F_3\equiv p=z;\hspace{6.5cm}\nonumber\\
F_2\equiv Q &\displaystyle =-\sqrt{3} Z\sqrt{ Z^2 + x^2 -\frac{\alpha _{m} }{2} x^4 +
\epsilon'_1 \left(z^2 -\mu^2 y^2 -\frac{\beta _{m} }{2} y^4 \right)+ \lambda _{m}} - x +\alpha _{m} x^{3};\nonumber\\
F_4\equiv q & \displaystyle =-\sqrt{3} z\sqrt{ Z^2 + x^2 -\frac{\alpha _{m} }{2} x^4 +
\epsilon'_1 \left(z^2 -\mu^2 y^2 -\frac{\beta _{m} }{2} y^4 \right)+ \lambda _{m}} +\mu^2 y +\beta _{m} y^{3}.
\end{eqnarray}
A corresponding canonical system of equations in standard notation has the following form:
\begin{equation} \label{DynSysPQ}
x'=P;\quad Z'=Q;\quad y'=p;\quad z'=q.
\end{equation}
In order the system of differential equations \eqref{DynSys} (or (\ref{DynSysPQ})) to have a real solution it is required an expression under radical to be non-negative:
\begin{equation} \label{GrindEQ__15_}
Z^2 + x^2 -\frac{\alpha _{m} }{2} x^4 +
\epsilon'_1 \left(z^2 -\mu^2 y^2 -\frac{\beta _{m} }{2} y^4 \right)+ \lambda _{m}\ge 0.
\end{equation}
\subsection{The Singular Points of the Dynamic System}
The singular points of the dynamic system are defined by the system of algebraic equations (see e.g.,  \cite{Bogoyav}, \cite{Bautin}):
\begin{equation} \label{GrindEQ__16_}
M:\quad F_i(x_1,\ldots,x_n)=0,\quad i=\overline{1,n}.
\end{equation}
Since according to (\ref{GrindEQ__13_}), (\ref{DynSysPQ}) and (\ref{GrindEQ__16_}) in the singular points of the dynamic system it is always
\begin{equation}\label{Z,z=0}
Z=0,\quad z=0,
\end{equation}
we obtain the following equation for finding the solutions:
\begin{eqnarray}\label{Eq_Sing_F}
x(1-\alpha_m x^2)=0;\\
y(1+\beta_m y^2)=0.
\end{eqnarray}
%
%%%%%%%%%%%%%%%%%%%%%%%%%%%%%%%%%%%%%%%%%%%%%%%%%%%%%%%%%%%%%%%%%%%%M0
\begin{enumerate} \label{sing_points}
\item $\mathbf{M_0}$: Thus, at any $\alpha _{m}$ and $\beta_m$ and $\lambda _{m} \ge 0$ the system of algebraic equations
\eqref{GrindEQ__16_}  has a trivial solution:
\begin{equation} \label{M0}
x=0;\quad Z=0;\quad y=0; \quad z=0 \Rightarrow M_{0}: (0,0,0,0).
\end{equation}
Substituting the solution \eqref{M0} into the condition \eqref{GrindEQ__15_}, we find a necessary condition for the solutions to be of a real type in the singular point:
\begin{equation} \label{ReM0}
(\ref{GrindEQ__15_})\rightarrow \lambda_m\ge 0.
\end{equation}
%%%%%%%%%%%%%%%%%%%%%%%%%%%%%%%%%%%%%%%%%%%%%%%%%%%%%%%%%%%%%%%%%%%%M01-M02
\item $\mathbf{M_{01},M_{02}}$: Next, at any $\alpha$ and $\beta<0$ we find two more conditions which are symmetric over $\phi$:
\begin{equation} \label{M01-M02}
x=0; Z=0;\quad y_\pm=\pm \frac{1}{\sqrt{-\beta_{m}}};\quad z=0\quad \Rightarrow M_{01}(0,0,|x_\pm|,0); \quad  M_{02}(0,0,-|x_\pm|,0).
\end{equation}
Substituting the solutions \eqref{M0} into the condition \eqref{GrindEQ__15_}, we find a necessary condition for the solutions to be of a real type in the singular points $M_{01},M_{02}$:
\begin{equation} \label{ReM01-M02}
(\ref{GrindEQ__15_})\rightarrow \frac{e'_1}{|\beta_m|}\left(\frac{1}{2}-\mu^2\right)+\lambda_m\ge 0.
\end{equation}
%
%%%%%%%%%%%%%%%%%%%%%%%%%%%%%%%%%%%%%%%%%%%%%%%%%%%%%%%%%%%%%%%%%%%%M10-M20
\item $\mathbf{M_{10},M_{20}}$: Further, at any $\beta$ and $\alpha>0$ we find two more conditions symmetrical over $\Phi$:
\begin{equation} \label{M10-M20}
x_\pm=\pm \frac{1}{\sqrt{\alpha_{m}}}; Z=0;\quad y=0;\quad z=0\quad \Rightarrow M_{10}(|x_\pm|,0,0,0); \quad  M_{20}(-|x_\pm|,0,0,0).
\end{equation}
Substituting the solutions \eqref{M10-M20} into the condition \eqref{GrindEQ__15_}, let us find a necessary condition of the solutions to be of a real type in the singular points $M_{10},M_{20}$:
\begin{equation} \label{ReM10-M20}
(\ref{GrindEQ__15_})\rightarrow \frac{2}{\alpha_m}+2\lambda_m - 1\ge 0.
\end{equation}
%
%%%%%%%%%%%%%%%%%%%%%%%%%%%%%%%%%%%%%%%%%%%%%%%%%%%%%%%%%%%%%%%%%%%%%%%%M12-M21;M11-M22
\item $\mathbf{M_{12},M_{21},M_{11}, M_{22}}$: Next, at $\beta<0$ and $\alpha>0$ we find 4 more conditions symmetrical over $\Phi$ and $\phi$:
\begin{eqnarray} \label{xpm,ypm}
x_\pm=\pm \frac{1}{\sqrt{\alpha_{m}}}; Z=0; & \displaystyle y_\pm=\pm \frac{1}{\sqrt{-\beta_{m}}};\quad z=0\quad \Rightarrow \\[12pt]
\label{M11-M12}
M_{11}(|x_\pm|,0,|y_\pm|,0); & \displaystyle M_{12}(|x_\pm|,0,-|y_\pm|,0); \\[12pt]
 \label{M21-M22}
M_{21}(-|x_\pm|,0,|y_pm|,0);& \displaystyle M_{22}(-|x_\pm|,0,-|y_\pm|,0).
\end{eqnarray}
Substituting the solutions \eqref{xpm,ypm} into the condition \eqref{GrindEQ__15_}, we obtain a necessary condition of the solutions to be of a real type in the singular points $M_{11},M_{12},M_{21},M_{22}$:
\begin{equation} \label{ReM11-M22}
(\ref{GrindEQ__15_})\rightarrow \frac{1}{2\alpha_m}+\frac{e'_1}{|\beta_m|}\left(\frac{1}{2}-\mu^2\right)+\lambda_m\ge 0.
\end{equation}
\end{enumerate}

Thus, the dynamic system (\ref{DynSys}) has 9 singular points. Let us investigate its character. The matrix of the dynamic system (\ref{DynSysPQ}), $A$, has the following form at $Z=z=0$:
\begin{equation}\label{A}
A=\left\|\frac{\partial F_i}{\partial x_k} \right\| =
\left\|
\displaystyle\begin{array}{cccc}
0 & 1 & 0 & 0\\
\displaystyle\frac{\partial Q}{\partial x} & \displaystyle\frac{\partial Q}{\partial Z} & 0 & 0\\
0 & 0 & 0 & 1\\
0 & 0 & \displaystyle\frac{\partial q}{\partial y} & \displaystyle\frac{\partial q}{\partial z}\\
\end{array}
\right\|_{Z=z=0} .
\end{equation}
The determinant of this matrix is defined by means of only partial derivatives  over dynamic variables $x,y$:
\begin{equation}\label{A}
\Delta(A)=\left|
\displaystyle\begin{array}{cc}
\displaystyle\frac{\partial Q}{\partial x} &  \displaystyle\frac{\partial Q}{\partial y} \\
\displaystyle\frac{\partial q}{\partial x} & \displaystyle\frac{\partial q}{\partial y} \\
\end{array}
\right| .
\end{equation}

\subsection{The characteristic Equation and the Qualitative Analysis in the Neighbourhood of a Singular Point}
Calculating the derivatives of the functions \eqref{GrindEQ__13_} in the null singular point \eqref{GrindEQ__16_} at $\lambda _{m} \ge 0$, we find a matrix of the system in this point:
\begin{equation}\label{A0}
A0\equiv A(M0)=\left\|
\displaystyle\begin{array}{cccc}
0 & 1 & 0 & 0\\
-1 & \displaystyle-\sqrt{3\pi\lambda_m} & 0 & 0\\
0 & 0 & 0 & 1\\
0 & 0 & \displaystyle\mu^2 & \displaystyle-\sqrt{3\pi\lambda_m}\\
\end{array}
\right\|.
\end{equation}
Its determinant is equal to:
\begin{equation}\label{Det0}
\Delta(A0)=-\mu^2\leqslant 0.
\end{equation}
Thus, we obtain a characteristic equation and its roots $k_i$
\begin{equation}\label{A0-k}
A0\equiv A(M0)=\left|
\displaystyle\begin{array}{cccc}
-k & 1 & 0 & 0\\
-1 & \displaystyle-\sqrt{3\pi\lambda_m}-k & 0 & 0\\
0 & 0 & -k & 1\\
0 & 0 & \displaystyle\mu^2 & \displaystyle-\sqrt{3\pi\lambda_m}-k\\
\end{array}
\right|=0.
\end{equation}
Calculating it, let us find the eigenvalues:
\begin{eqnarray}\label{k(M0)}
k_1(M0)=-\xi+\sqrt{\sigma};\; k_2(M0)=-\xi-\sqrt{\sigma};k_3(M0)=-\xi+\zeta;\; k_4(M0)=-\xi-\zeta;\; \\
\xi^2\equiv\frac{3\pi\lambda_m}{4}>0;\; \zeta^2\equiv \xi^2+\mu^2>\xi^2;\; \sigma=\xi^2-1<\xi^2.
\end{eqnarray}
Thus,
\begin{equation}\label{signum_kM0}
\mathrm{Re}(k_1)<0; \quad \mathrm{Re}(k_2)<0;\quad k_3>0;\quad k_4<0;\quad .
\end{equation}
Therefore according to the qualitative theory of the dynamic systems \cite{Bogoyav} a non-degenerated point $M0$ is a \emph{saddle point}, where all the trajectories which are outbound from this point, lay in a single-dimensional manifold $W^1_u$, and all the inbound to this point trajectories lay in a 3-dimensional invariant manifold $W^3_s$. In the neighbourhood of the singular points $M(x_i^0)$ (\ref{GrindEQ__16_}) the asymptotic solutions of the system (\ref{standart_sys}) have a next form:
\begin{equation}\label{asympt_sol}
x_i(\tau)=x_i^0+\mathrm{Re}\left(\sum\limits_{j=1}^n C_j u^j_i \mathrm{e}^{k_j \tau}\right),
\end{equation}
where $u^j_i$ is an eigenvector of the matrix $A(M0)$, corresponding to the eigenvalue $k^j$. Let us find these vectors:
\begin{eqnarray}\label{u0}
\mathbf{u}_1(M0)= [1,-\xi+\sqrt{\sigma},0,0]; & \mathbf{u}_2(M0)= [1,-\xi-\sqrt{\sigma},0,0];\nonumber\\
\mathbf{u}_3(M0)= [0,0,1,-\xi+\zeta]; & \mathbf{u}_4(M0)=  [0,0,1,-\xi-\zeta].
\end{eqnarray}
Therefore, the phase trajectories outcome from the point $M0$ tangentially to the vector $\mathbf{u}_3(M0)$, and income to this point at $t\to+\infty$ in a tangential 3-dimensional space $V_3(\mathbf{u}_1,\mathbf{u}_2,\mathbf{u}_4)$. Thus, a classical scalar field disappears in this singular point.

Let us notice that at significantly small values of the cosmological constant
\begin{equation}\label{lambda_to_0}
\xi^2<1\Rightarrow \lambda_m<\frac{4}{3\pi}
\end{equation}
the eigenvalues $k_1(M0),k_2(M0)$ (\ref{k(M0)}) become complex conjugated with positive real parts. In the absence of phantom scalar field, when the dynamic system becomes 2-dimensional, the null point $M0$ is an attractive focus for the classical field (at $\lambda\equiv0$), or an attractive center at $\lambda>0$   (see \cite{Ignat_16_1_stfi}, \cite{Ignat_16_2_stfi}, \cite{Ignat_17_1_GC}).
A presence of a weak phantom field radically changes the situation: the point $M0$ becomes a saddle one. In the neighbourhood of this point the phase trajectories have the following asymptotic in accordance with
(\ref{asympt_sol}):
\begin{equation}\label{asympt_sol_0}
x_i(\tau)=\sim u^j_i \mathrm{e}^{\sqrt{\sigma}\tau\pm ib^2\tau};\quad (\sigma>0),
\end{equation}
i.e. they repulse from this singular point in an oscillating mode in the direction $\{\Phi,Z\}$.

Fig. \ref{ris_fasef=0} and \ref{ris_fasePhi=0} show the projections of the phase trajectories of the dynamic system (\ref{DynSys}) in the planes
$\{\Phi,Z\}$ è $\{\phi,z\}$ for a phantom field with attraction in the case when one of the scalar fields is disabled. Results in this case, as expected,  coincide with the cited above results of the previous works. All the graphs investigate a single case of $\mu=1$, i.e, $\mathfrak{m}=m$.

\TwoFig{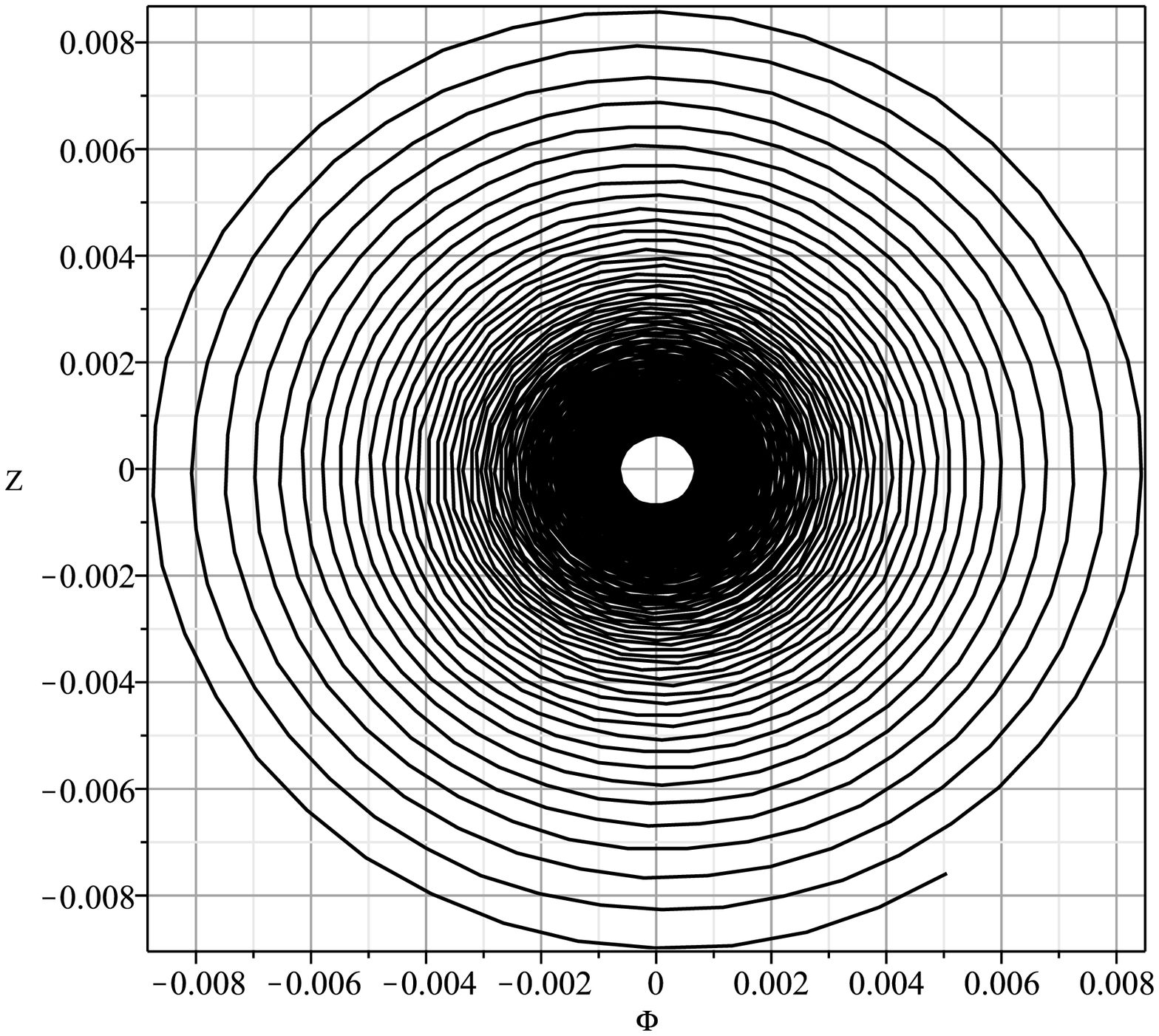}{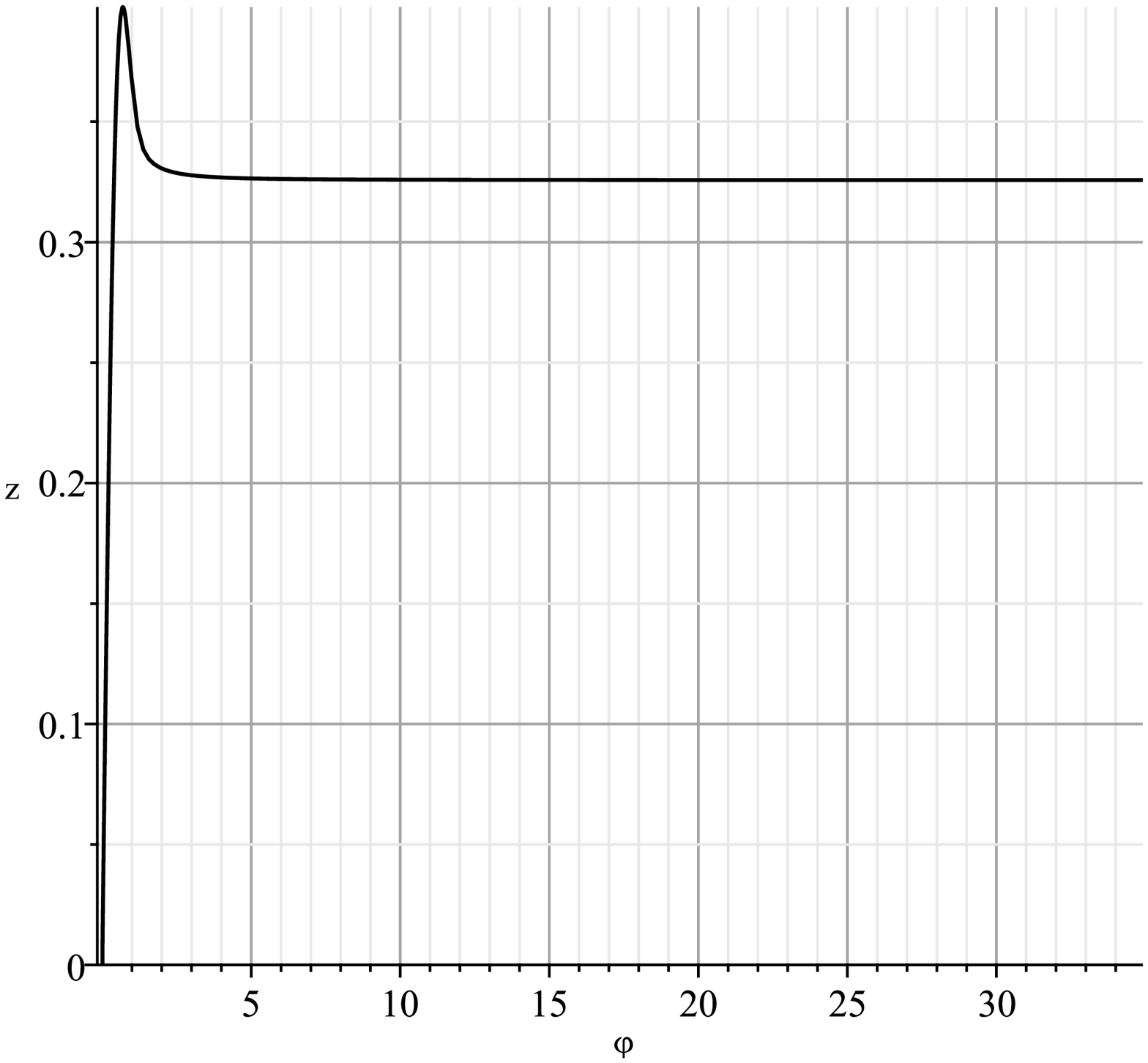}{\label{ris_fasef=0}The phase trajectory of the dynamic system without a self-action at absence of the phantom field ($\phi(-10)=0;\ z(=10)=0$) in the plane $\{\Phi,Z\}$; $\Phi(-10)=10,Z(-10)=0$; $\lambda=0$.}{\label{ris_fasePhi=0}The phase trajectory of the dynamic system without a self-action at absence of the classical field  ($\Phi(\tau_0)=0;\ Z(\tau_0)=0$) in the plane $\{\phi,z\}$; $\phi(-10)=0.01,z(-10)=0$; $\lambda=0$, $\epsilon1_1=-1$.}

However, at appearance of even a very weak phantom field the behaviour of the classical field in the neighbourhood of the finite point $M0$ radically changes though in a large scale it just slightly differs from the phase trajectory of a single field (Fig. \ref{ris_FasePhiZ_comb}, \ref{ris_FasePhiZ_comb_end}). The behaviour of the phantom field almost does not depend on a classical field (\ref{ris_Fasephzphi_comb}). Fig. \ref{ris_FasePhiZphi_comb_end} illustrates a 3-dimensional projection of the phase trajectory of this system.

\TwoFig{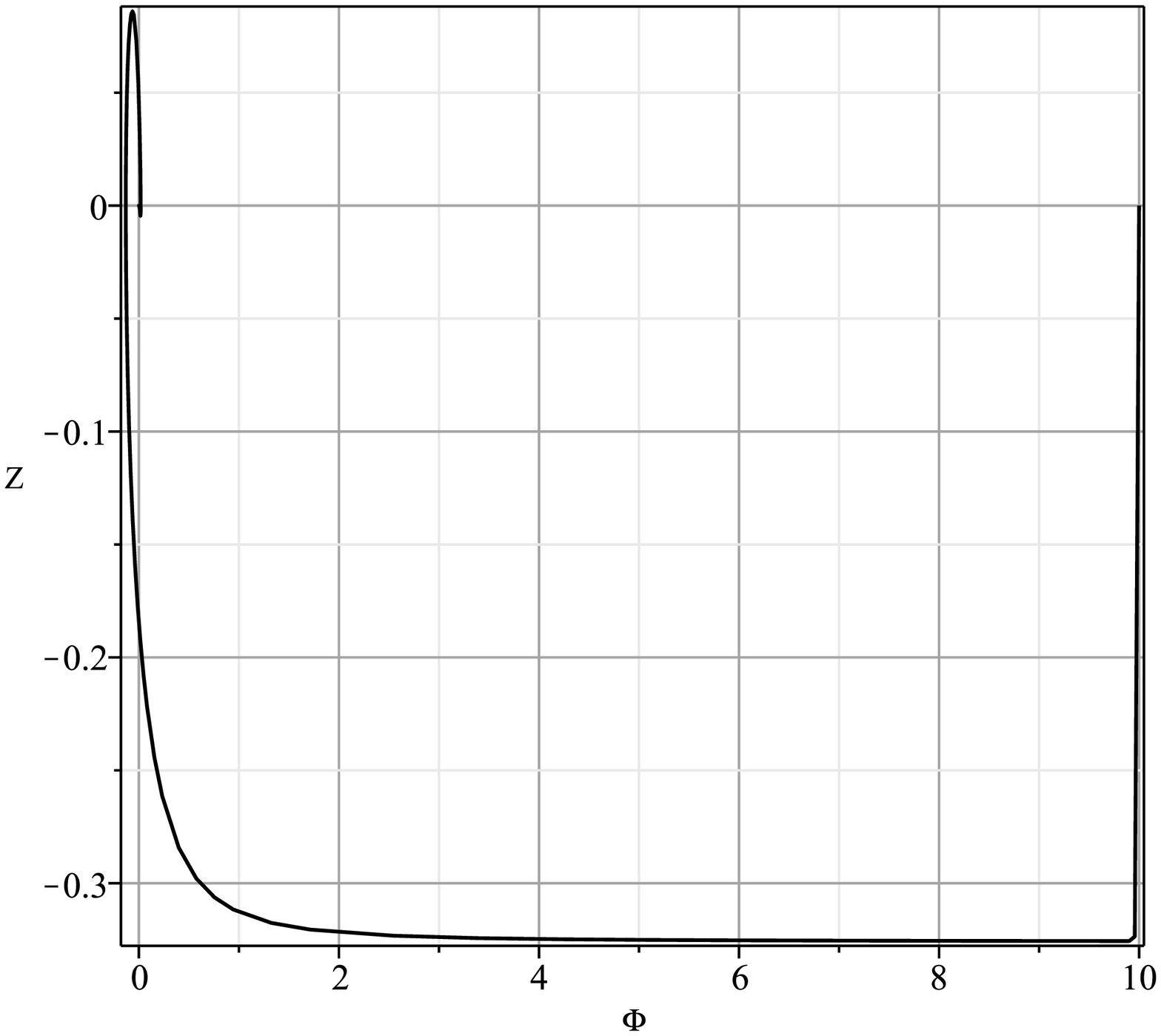}{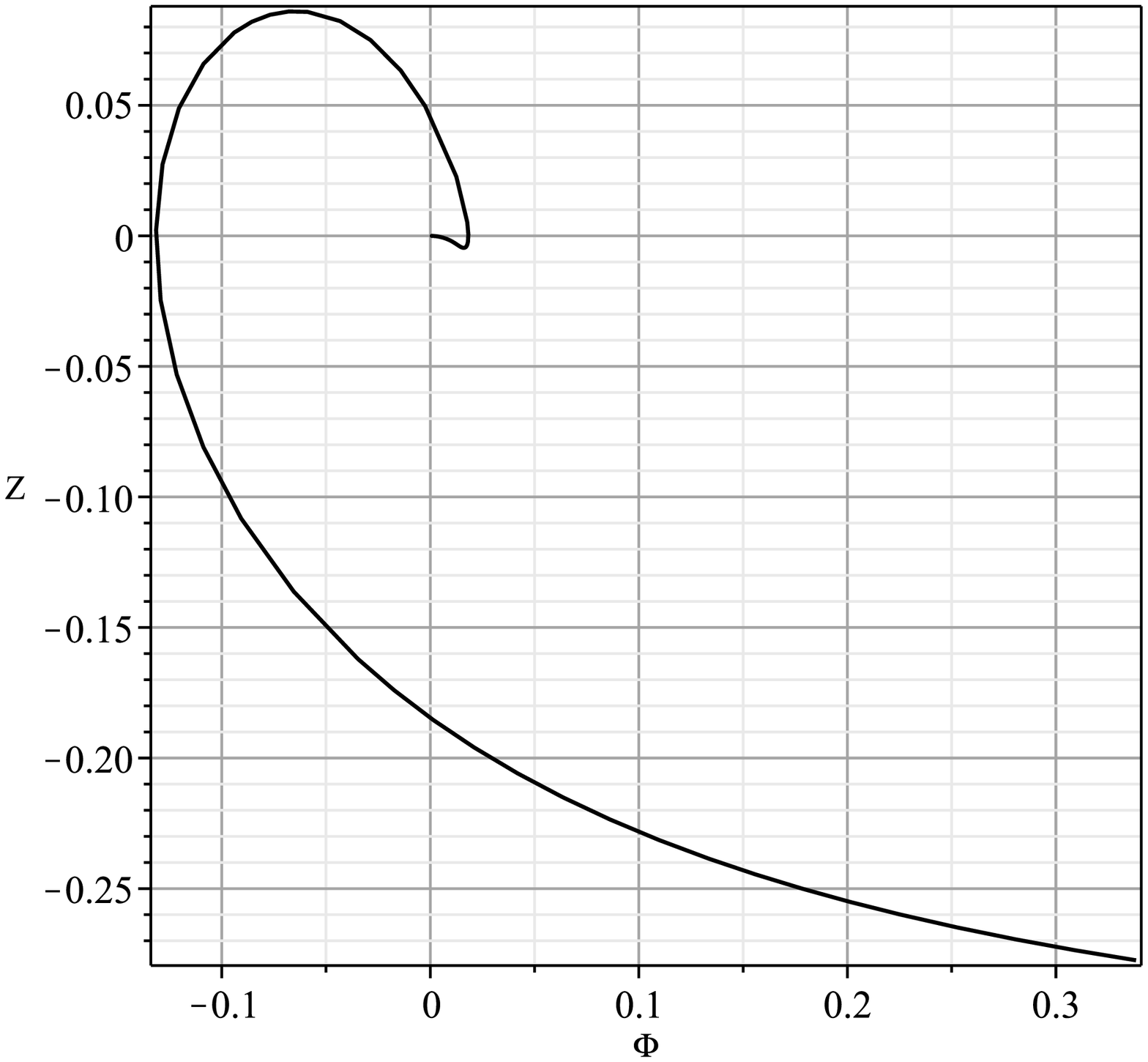}{\label{ris_FasePhiZ_comb}A phase trajectory of the dynamic system without a self-action at presence of a weak phantom field with attraction ($\phi(-10)=0.01;\ z(-10)=0$) in the plane $\{\Phi,Z\}$; $\Phi(-10)=10,Z(-10)=0$ in a large scale; $\lambda=0$.}{\label{ris_FasePhiZ_comb_end}The phase trajectory of the dynamic system without a self-action at presence of a weak phantom field with attraction  ($\Phi(\tau_0)=10;\ Z(\tau_0)=0$) in the plane $\{\Phi,Z\}$ on a final stage; $\phi(-10)=0.01,z(-10)=0$; $\lambda=0$, $\epsilon1_1=-1$.}

\TwoFig{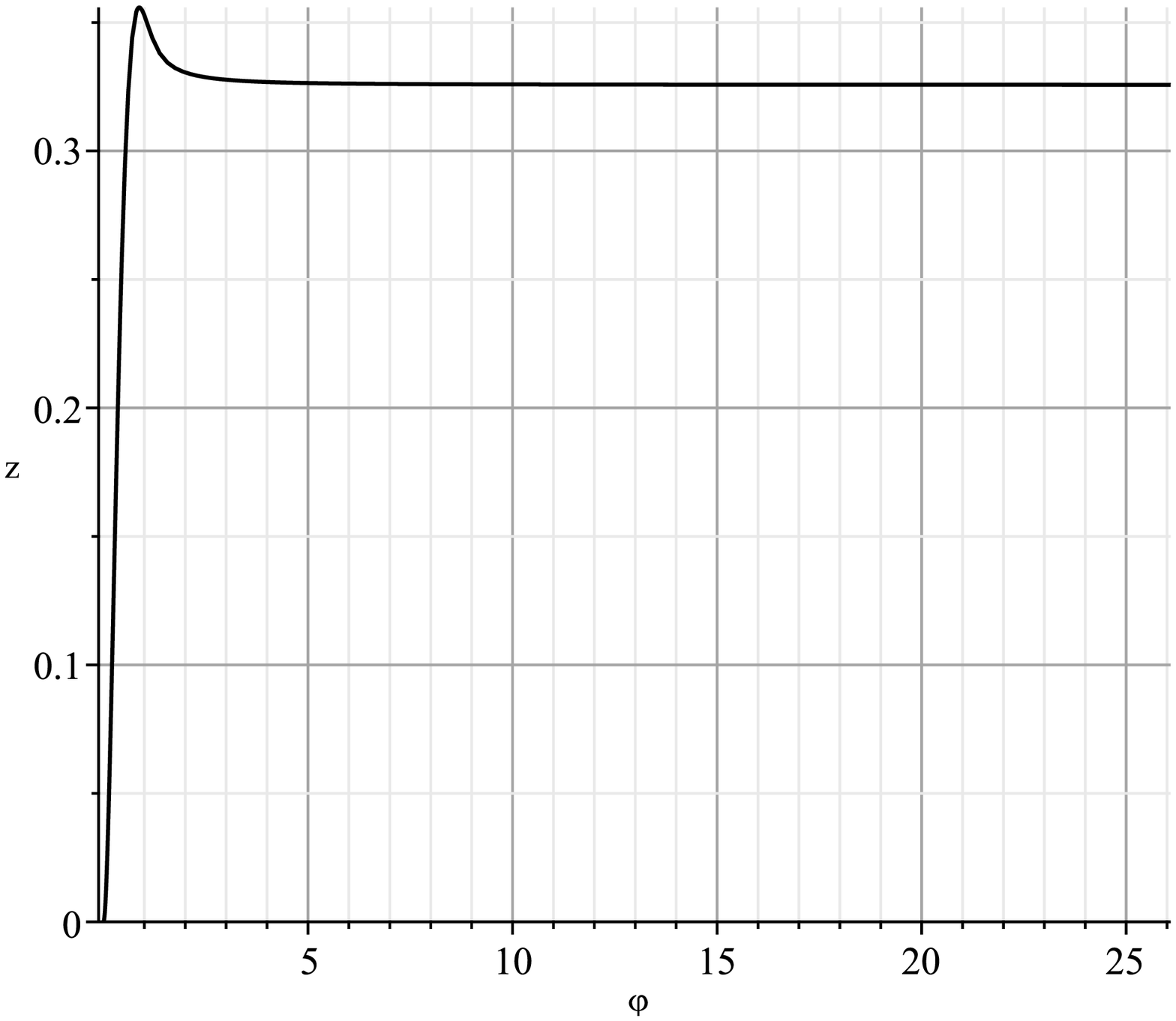}{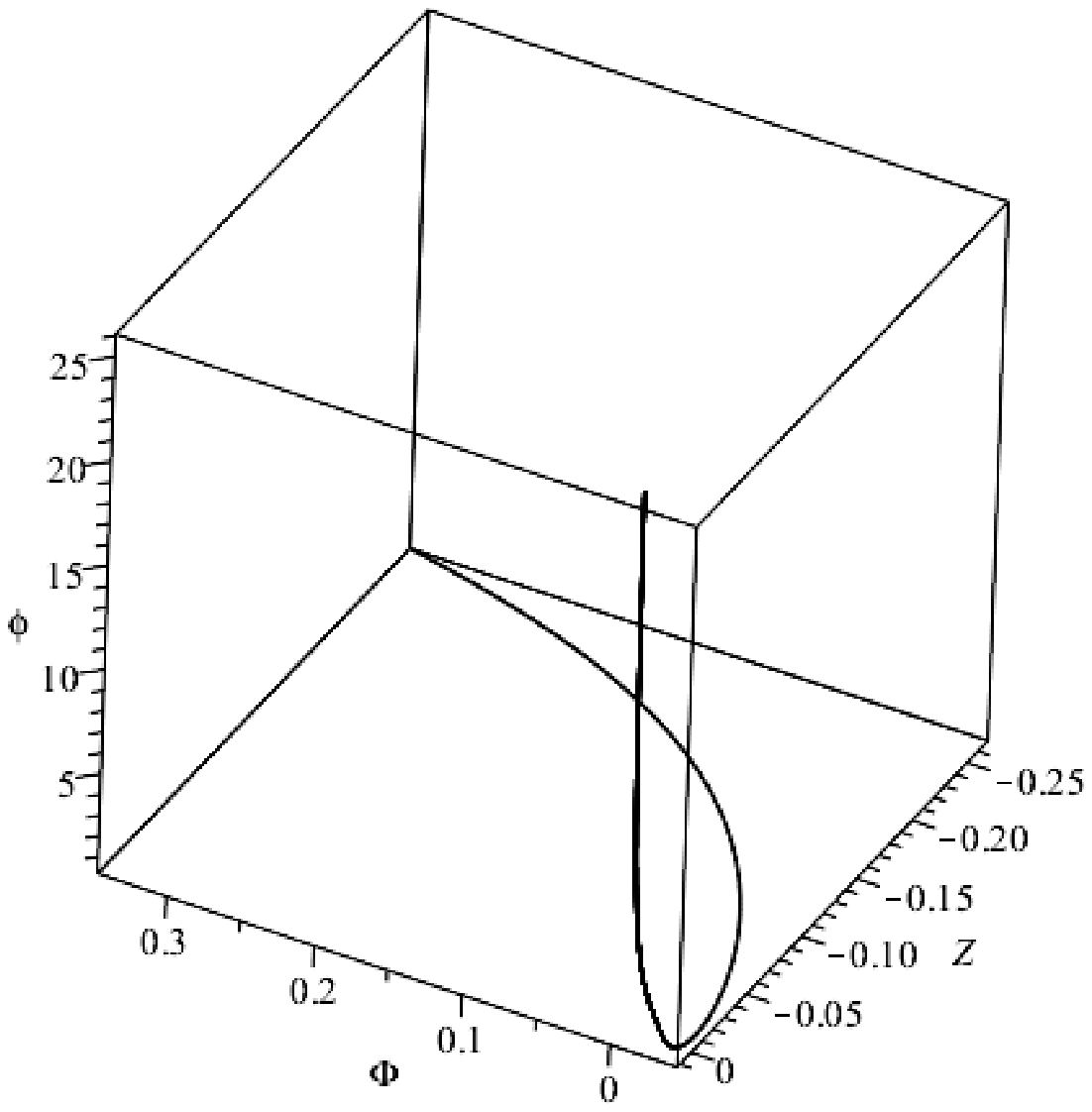}{\label{ris_Fasephzphi_comb}A phase trajectory of the dynamic system without a self-action at presence of a weak phantom field with attraction ($\phi(-10)=0.01;\ z(-10)=0$) in the plane $\{\phi,z\}$; $\Phi(-10)=10,Z(-10)=0$ in a large scale; $\lambda=0$.}{\label{ris_FasePhiZphi_comb_end}A 3-dimensional projection of the phase trajectory of the dynamic system without a self-action of a classical field ($\Phi(-10)=10;\ Z(-10)=0$; $\phi(-10)=0.01;\ z(-10)=0$)) in a subspace $\{\Phi,Z,\phi\}$; $\lambda=0$, $\epsilon1_1=-1$.}

\TwoFig{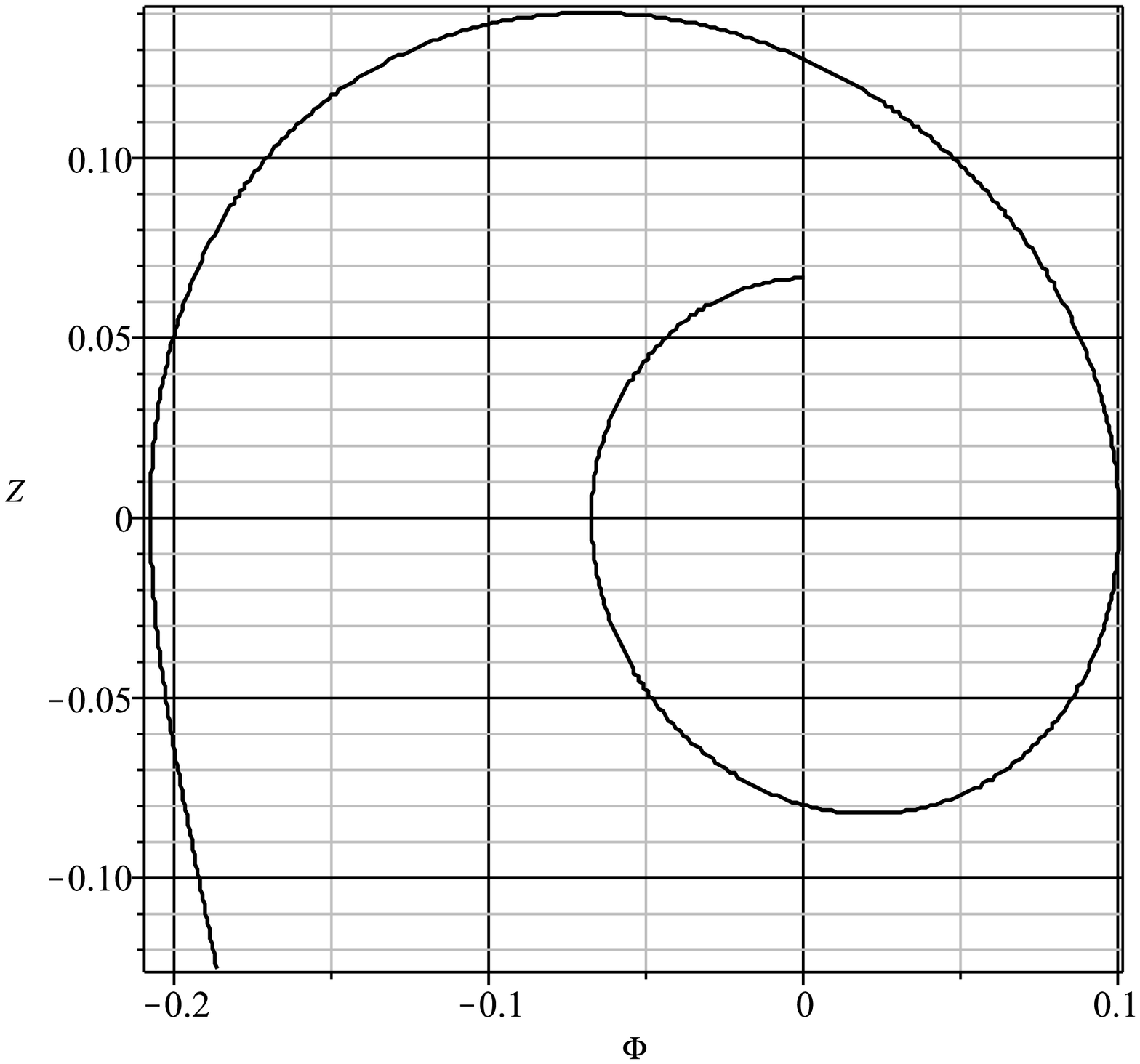}{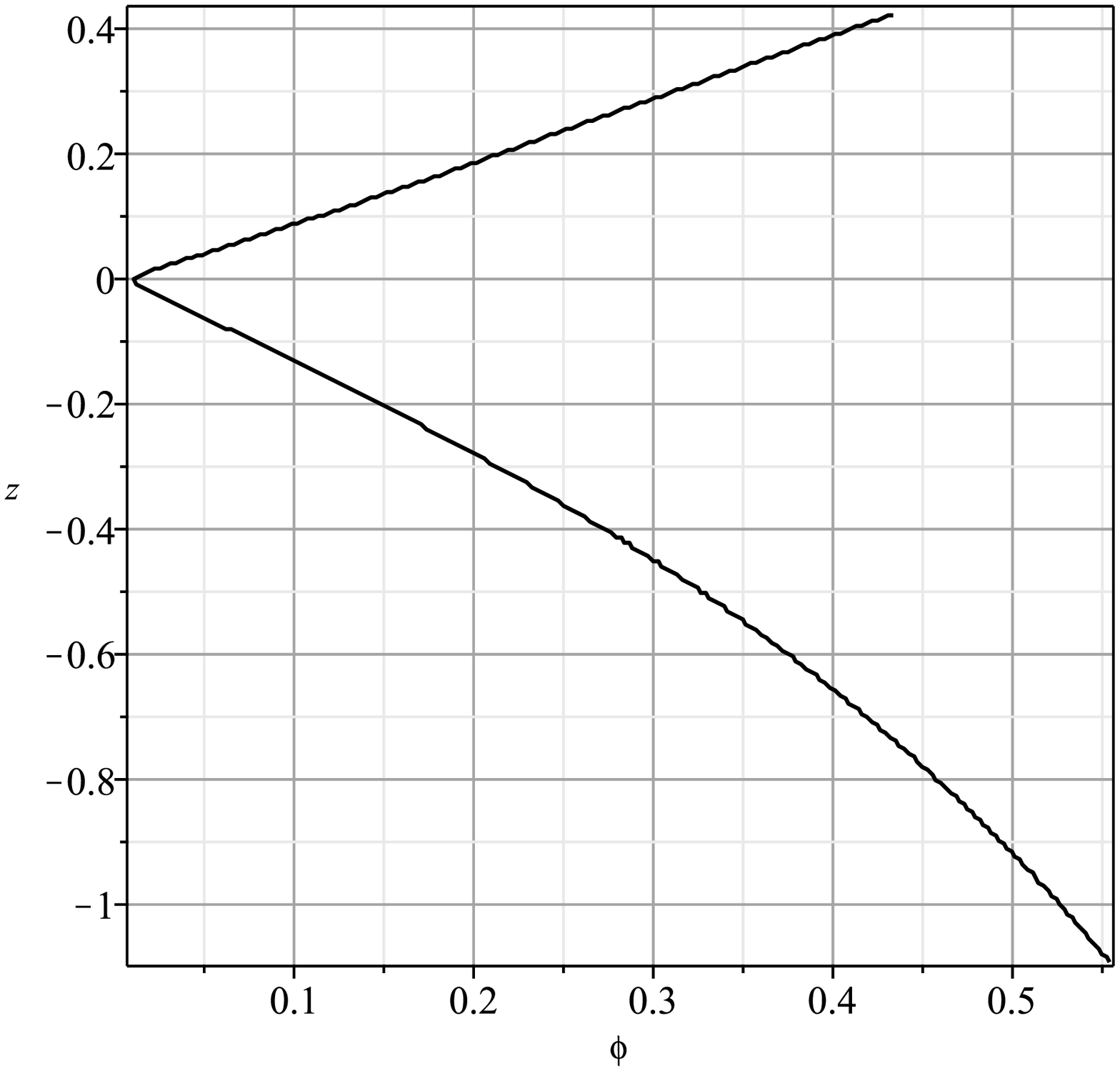}{\label{ris_FasePhiZ_comb}A phase trajectory of the dynamic system with a self-action ($\alpha=0.1; \beta=-0.1$) in case of presence of a weak phantom field with repulsion ($\phi(0)=0.1;\ z(0)=0$) in the plane $\{\Phi,Z\}$; $\Phi(0)=0.1,Z(0)=0$; $\lambda=0$, $\epsilon1_1=1$..}{\label{ris_FasePhiZ_comb_end}A phase trajectory of the dynamic system without a self-action at presence of a weak phantom field with repulsion ($\Phi(\tau_0)=10;\ Z(\tau_0)=0$) in the plane $\{\Phi,Z\}$ on a final stage; $\phi(-10)=0.01,z(-10)=0$; $\lambda=0$, $\epsilon1_1=1$.}

\TwoFig{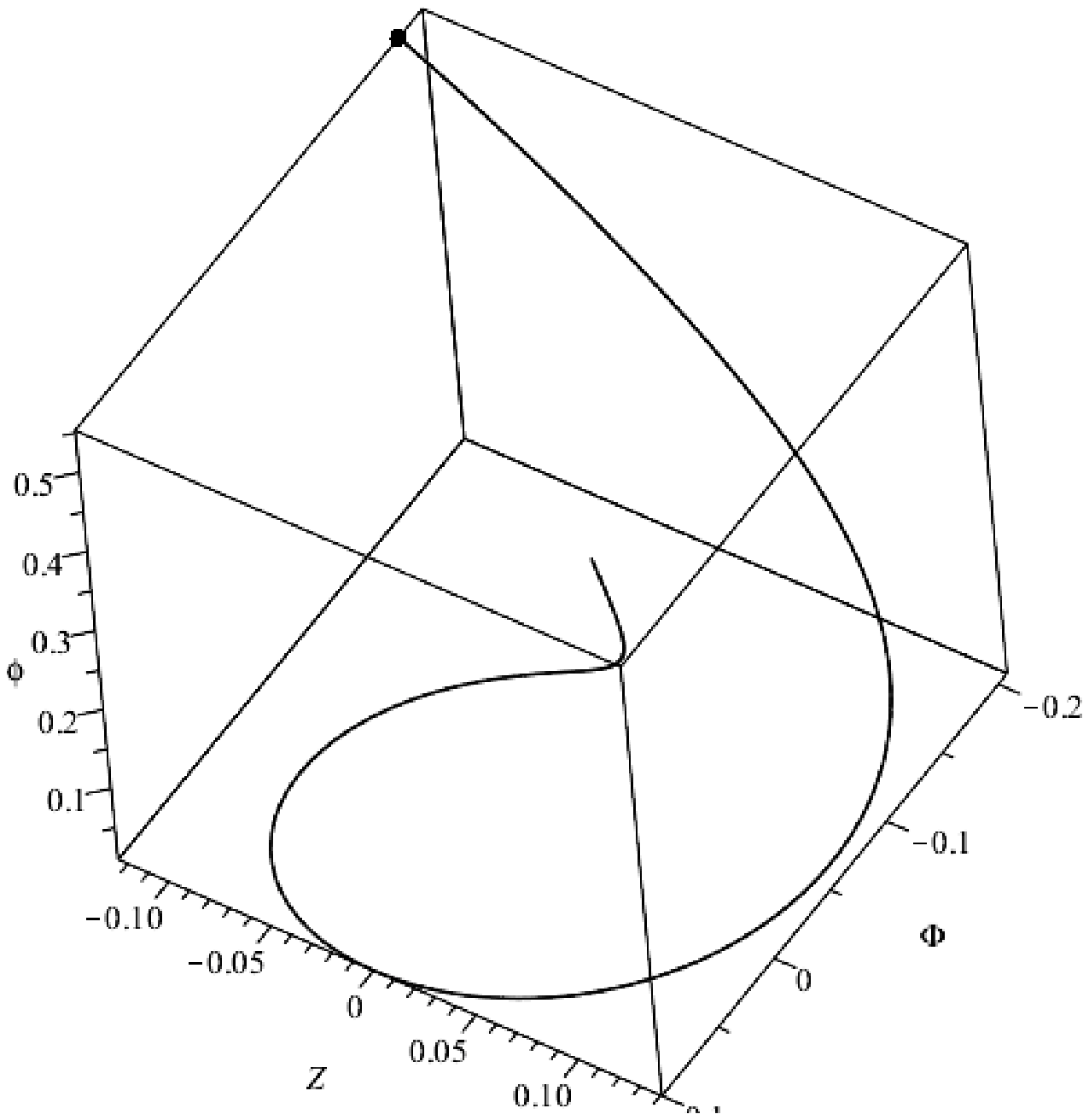}{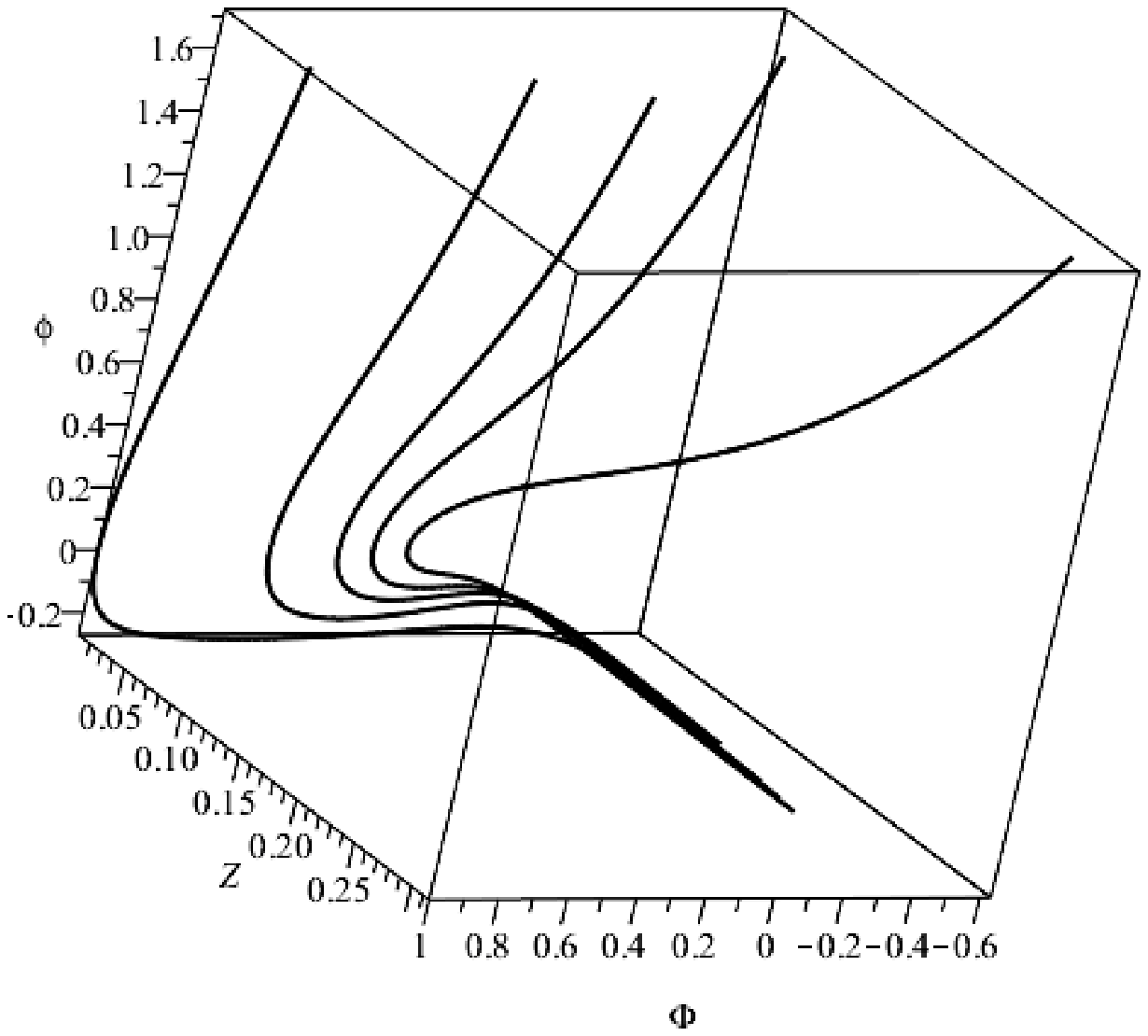}{\label{ris_FasePhiZphi_comb_self}A 3-dimensional projection of the phase trajectory of the dynamic system in presence of a weak phantom field with repulsion ($\phi(0)=0.01;\ z(0)=0$) in a subspace $\{\Phi,Z,\phi\}$; $\Phi(0)=0.1,Z(0)=0$; $\alpha=0.1;\beta=-0.1$; $\lambda=0$.}{\label{ris_FasePhiZphi_comb_self_gr}A 3-dimensional projection of the phase trajectory of the dynamic system in presence of a weak phantom field with repulsion ($\phi(0)=0.1;\ z(0)=0$) in a subspace $\{\Phi,Z,\phi\}$; $\Phi(0)=0.1; 0.2; 0.3; 0.5; 1,Z(0)=0$ in a large scale; $\lambda=0$; $\alpha=0.1;\beta=-0.1; \epsilon'_1=1$.}

\TwoFig{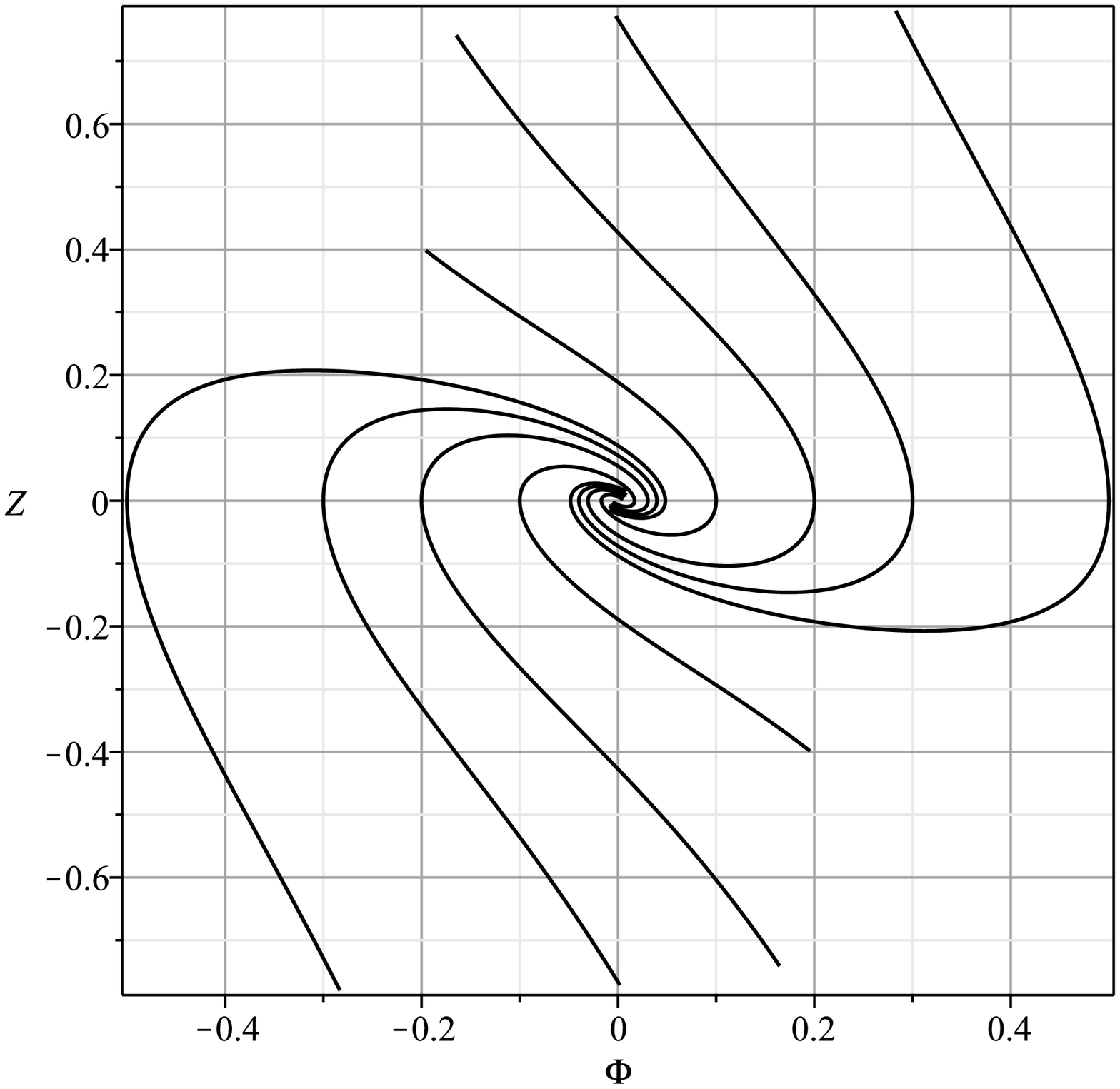}{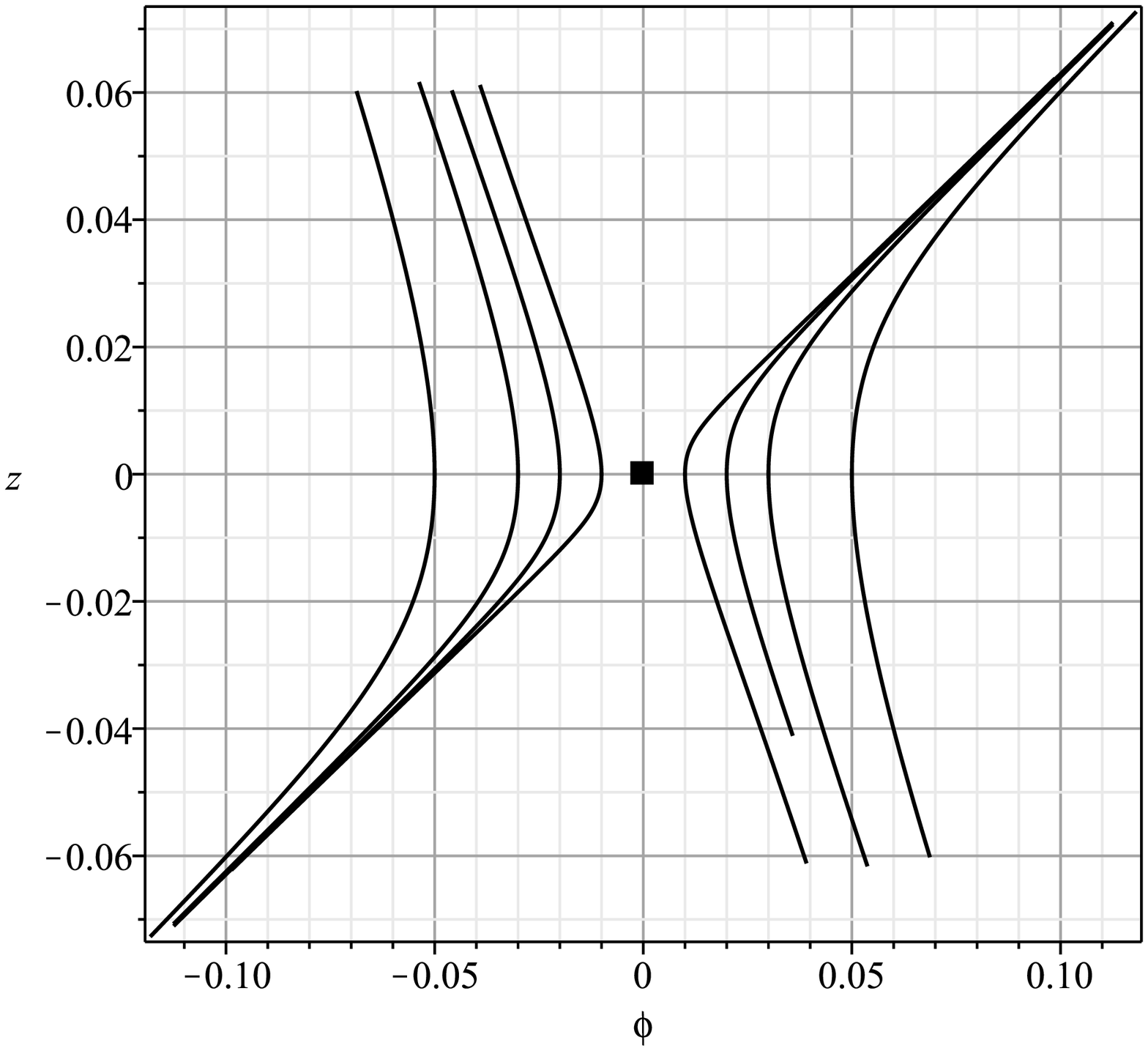}{\label{ris_Gr_Clas}Phase trajectories of the dynamic system with a self-action in presence of a weak phantom field with repulsion in the plane $\{\Phi,Z\}$;  $\lambda=0.1;\alpha=0.1;\beta=-0.1$; ($\Phi(0)=0.1;0.2;0.3;0.5$; $\Phi(0)=-0.1;-0.2;-0.3;-0.5$; $Z(0)=0$). $\phi(0)=0.01;z(0)=0$.}{\label{ris_Grfantom}Phase trajectories of the dynamic system with a self-action in presence of a weak phantom field with repulsion in the plane $\{\Phi,Z\}$; $\Phi(0)=0.1,Z(0)=0$; $\lambda=0.1;\alpha=0.1;\beta=-0.1$; ($\phi(0)=0.01;0.02;0.03;0.05$; $\phi(0)=-0.01;-0.02;-0.03;-0.05$ $z(0)=0$).}

\subsection{The Characteristic Equation and the Qualitative Analysis in the Neighbourhood of Singular Points $M_{01},M_{02}$}
Calculating the derivatives of the functions \eqref{GrindEQ__13_} in points (\ref{M01-M02}) at $\lambda _{m} \ge 0$, we find in these papers the matrix of the system:
\begin{equation}\label{A01}
A01\equiv A(M_{01})=A(M_{02})=\left\|
\displaystyle\begin{array}{cccc}
0 & 1 & 0 & 0\\
-1 & \displaystyle-2\rho & 0 & 0\\
0 & 0 & 0 & 1\\
0 & 0 & \displaystyle -2\mu^2 & \displaystyle -2\rho\\
\end{array}
\right\|,
\end{equation}
where it is
\begin{equation}\label{rho}
\rho=\sqrt{\frac{3\pi}{4}\left(\frac{e'_1\mu^4}{\beta}+2\lambda_m\right)},
\end{equation}
-- in consequence of (\ref{ReM01-M02}) its determinant is greater than null:
\begin{equation}\label{Det0}
\Delta(A01)=2\mu^2\geqslant 0.
\end{equation}
Thus, we obtain the roots $k_i$ of the characteristic equation:

\begin{eqnarray}\label{k(M01)}
k_1(M01)=-\rho+\sqrt{\nu};\; k_2(M01)=-\rho-\sqrt{\nu};k_3(M01)=-\rho+\delta;\; k_4(M01)=-\rho-\delta; \\
\delta^2\equiv \rho^2-2\mu^2<\rho^2;\; \nu=\rho^2-1<\rho^2.
\end{eqnarray}
\begin{equation}\label{signum_kM0}
\mathrm{Re}(k_1)<0; \quad \mathrm{Re}(k_2)<0;\quad k_3>0;\quad k_4<0;\quad .
\end{equation}
Thus, the real parts of all the eigenvalues are negative i.e., \emph{points $M_{01},M_{02}$ are attracting points}. In the neighbourhood of these singular points asymptotic solutions of the system (\ref{DynSys}) can either smoothly approach to these points or wind round them depending on signs of $\nu,\delta$ .

\subsection{The Characteristic Equation and the Qualitative Analysis in the Neighbourhood of Singular Points $M_{10},M_{20}$}
Calculating the derivatives of the functions \eqref{GrindEQ__13_} in points (\ref{M01-M02}) at $\lambda _{m} \ge 0$, we find the matrix of the system in these points:
\begin{equation}\label{A01}
A01\equiv A(M_{01})=A(M_{02})=\left\|
\displaystyle\begin{array}{cccc}
0 & 1 & 0 & 0\\
2 & \displaystyle-2\chi & 0 & 0\\
0 & 0 & 0 & 1\\
0 & 0 & \displaystyle -2\mu^2 & \displaystyle -2\chi\\
\end{array}
\right\|,
\end{equation}
where
\begin{equation}\label{rho}
\chi=\sqrt{\frac{3\pi}{4}}\sqrt{2\lambda_m-1+\frac{2}{\alpha}},
\end{equation}
-- in consequence of (\ref{ReM10-M20}) the radicand is greater than null, the determinant of the matrix is also greater than null:
\begin{equation}\label{Det0}
\Delta(A01)=2\mu^2\geqslant 0.
\end{equation}
Thus, we obtain the roots $k_i$ of the characteristic equation:

\begin{eqnarray}\label{k(M01)}
k_1(M10)=-\chi+\sqrt{\omega};\; k_2(M10)=-\chi-\sqrt{\omega};k_3(M10)=-\chi+\sqrt{\eta};\; k_4(M01)=-\chi-sqrt{\eta}; \\
\omega\equiv \chi^2+\mu^2>\chi^2;\; \eta=\chi^2+2>\rho^2.
\end{eqnarray}
Thus. it is
\begin{equation}\label{signum_kM0}
k_1<0; \quad k_2<0;\quad k_3>0;\quad k_4<0.
\end{equation}
Therefore, the eigenvalues are pairwise-opposite in terms of a sign, i.e. \emph{the points $M_{10},M_{20}$ are saddle ones}, and all the trajectories that are outbound from this point lie in a 2-dimensional manifold $W^2_u$, while all the trajectories which are inbound to this point at $\tau\to+\infty$ income to this point over a 2-dimensional manifold $W^2_s$.

\subsection{The Characteristic Equation and the Qualitative Analysis in the Neighbourhood of the Singular Points $M_{11},M_{22}, M_{12}, M_{21}$}
Similarly, we can show that the points $M_{11},M_{22}, M_{12}, M_{21}$ as well as the point $M_0$, are \emph{saddle singular points}, and all the trajectories that are outbound from these points lie in single-dimensional manifolds $W^1_u$, and all the trajectories inbound to these points at $\tau\to+\infty$
income to this point over a 3-dimensional invariant manifold $W^3_s$.

Fig. \ref{SingPoints} illustrates a qualitative picture of situation of the dynamic system's singular points in the plane $\{\Phi,Z\}$.

\Fig{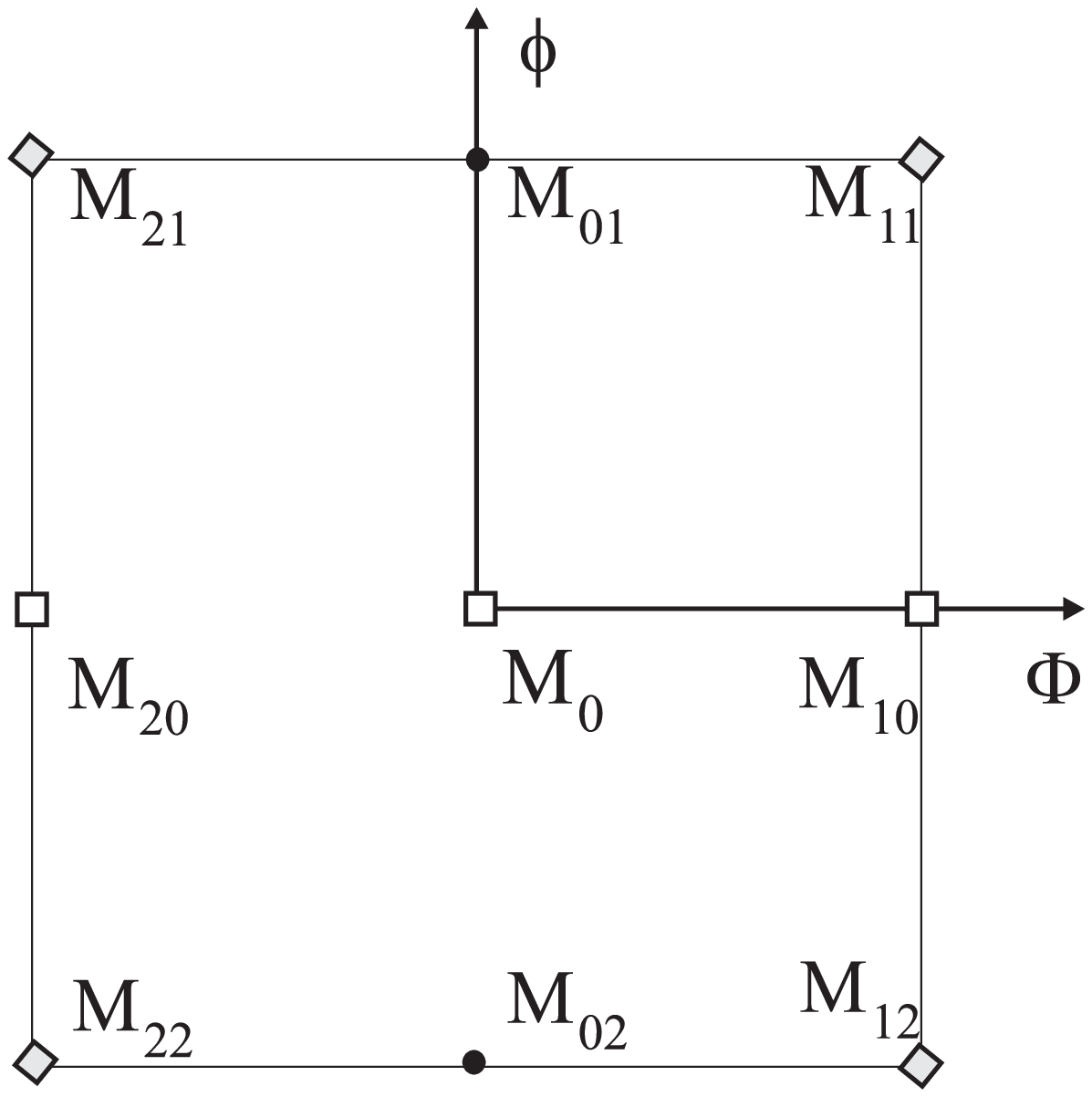}{8}{\label{SingPoints}The singular points of the dynamic system (\ref{DynSys}) in the phase space $Z=z=0$. Squares are used to mark saddle points with invariant manifolds of a type $W^2_s\times W^2_u$, while rhombuses are to mark saddle points with invariant manifolds of type $W^3_s\times W^31_u$, and black circles designate attractive points.}

\section{The Conclusion}
The performed analysis showed, first of all, a great diversity of behaviours of the cosmological system based on asymmetric scalar doublet, and second, a significant influence of a phantom scalar field on the dynamics of a classical scalar field even in a minimal model of interaction. In the same time, a classical scalar field has almost no influence on the dynamics of a phantom scalar field. This property can be used in the cosmological models for regulating their behaviour with a help of small phantom fields.

%%%%%%%%%%%%%%%%%%%%%%%%%%%%%%%%%%%%%%%%%%%%%%%%%%%%%%%%%%%%%%%%%%%%%%%%
%% Áèáëèîãðàôèÿ %%%%%%%%%%%%%%%%%%%%%%%%%%%%%%%%%%%%%%%%%%%%%%%%%%%%%%%%
%%%%%%%%%%%%%%%%%%%%%%%%%%%%%%%%%%%%%%%%%%%%%%%%%%%%%%%%%%%%%%%%%%%%%%%%
\vspace{20pt}

\small

\makeatletter
\@addtoreset{equation}{section}
\@addtoreset{footnote}{section}
\renewcommand{\section}{\@startsection{section}{1}{0pt}{1.3ex
plus 1ex minus 1ex}{1.3ex plus .1ex}{}}

{ %\scriptsize

\renewcommand{\refname}{{\rm\centerline{ÑÏÈÑÎÊ ËÈÒÅÐÀÒÓÐÛ}}}

\end{document}